# Metal-insulator transition and electrically-driven memristive characteristics of SmNiO$_3$ thin films

Sieu D. Ha*, Gulgun H. Aydogdu*, and Shriram Ramanathan

*School of Engineering and Applied Sciences, Harvard University, Cambridge, MA 02138*

*Equally contributing authors

**Abstract**

The correlated oxide SmNiO$_3$ (SNO) exhibits an insulator to metal transition (MIT) at 130 °C in bulk form. We report on synthesis and electron transport in SNO films deposited on LaAlO$_3$ (LAO) and Si single crystals. X-ray diffraction studies show that compressively strained single-phase SNO grows epitaxially on LAO while on Si, mixed oxide phases are observed. MIT is observed in resistance-temperature measurements in films grown on both substrates, with charge transport in-plane for LAO/SNO films and out-of-plane for Si/SNO films. Electrically-driven memristive behavior is realized in LAO/SNO films, suggesting that SNO may be relevant for neuromorphic devices.

The rare earth nickelates have been of particular interest since the discovery of a temperature-driven metal-insulator transition (MIT) in several lanthanide compounds (LnNiO$_3$).[1] The transition temperature ($T_{MIT}$), generally defined as the temperature at which $d\rho/dT$ changes sign (where $\rho$ is the resistivity), systematically increases with decreasing rare earth cation radius, which influences the electronic band structure and therefore $T_{MIT}$ through the Ni-O bond distance and Ni-O-Ni bond angle.[2] Such oxides that exhibit an MIT make candidate components for electronics such as neural circuits[3] and nonvolatile memory[4] that require fast switching speed and scalability. SmNiO$_3$ (SNO), with a transition at 130 °C, is the first rare earth nickelate in the series with $T_{MIT}$ above room temperature. This has important implications for integration of functional oxides into CMOS technology given that typical processor junction temperatures are ~90 °C. Thus, SNO is a candidate material for investigation into oxide electronics involving metal-insulator transitions.

Among the applications of MIT oxides are memristive systems that can be used for neuromorphic learning circuits, which are capable of analog computation that can be complex for digital circuits.[3] In this work, we present electrical measurements with memristor characteristics in SNO thin film devices.[5] In addition, in order to exploit the possibilities of SNO, it is necessary not just to fabricate devices with an in-plane MIT,[6] but also out-of-plane MIT. Such vertical device geometries may be relevant for applications such as switchable or reconfigurable interconnects and three-dimensional integrated circuits.[7] In this manuscript, we demonstrate synthesis of SNO thin films by RF sputtering as well as MIT characteristics in in-plane and out-of-plane devices.

SNO thin films were deposited on LaAlO$_3$ (001) (LAO; nominal thickness ~80 and 200 nm; deposition rate calibrated by x-ray reflectivity) for in-plane devices and highly doped Si

(100) (nominal thickness ~120 nm) for out-of-plane devices by RF magnetron sputtering with 200 W target power using an SNO target of 99.99% purity (AJA International, Inc.) under 0.01 mbar total pressure at 650 °C. Reports of RF sputtering deposition of SNO are scarce possibly because high pressure $O_2$ annealing may be required to stabilize the $Ni^{3+}$ valence state, as for $NdNiO_3$.[8] The crystal structure, phase analysis, and epitaxial relationship between film and substrate were examined with x-ray diffraction (XRD). $2\theta$-$\omega$ scans, reciprocal space mapping (RSM), and $\varphi$ scans were performed with a four-circle Bruker D8 Discover diffractometer equipped with Göbel mirror, 4-bounce 022 Ge channel-cut monochromator, and PANalytical X'Celerator diffractometer. X-ray photoelectron spectroscopy (XPS) measurements were carried out with an ESCA SSX-100 spectrometer using Al K$\alpha$ radiation and constant pass energy of 100 eV. For in-plane resistance-temperature (*R-T*) measurements, 2-terminal Cr/Au devices were patterned by photolithography with 20 μm channel length and 180-240 μm channel width. For out-of-plane measurements, top Cr/Au contacts with areas of 0.01-0.25 mm$^2$ were deposited through a shadow mask. Ag paste was used as bottom contact to the conducting Si (100) substrate ($\rho$ = 0.002-0.005 Ω·cm).

The $2\theta$-$\omega$ scan of as-deposited films shown in Fig. 1(a) indicates that single phase SNO films are epitaxially grown c-axis oriented on LAO substrates. As for the in-plane film orientation, $\varphi$ scan (Fig. 1(a) inset) performed on the (022) plane of SNO confirms that films on LAO show four-fold symmetry. The lattice parameter of LAO (pseudocubic parameter ~0.379 nm) is smaller than the lattice parameter of SNO (~0.3796 nm for bulk SNO), producing a compression of ~-0.15 % in the film. The strain states in the films were examined using RSM. Representative RSM images for the 103 asymmetric reflections of LAO/SNO are shown for the 80 nm and 200 nm films in Figs. 1(c) and 1(d). The ordinate (abscissa) of these figures corresponds to the inverse out-of-plane (in-plane) lattice parameter. The in-plane components of wave vectors pertaining to 103 reflection of both SNO films and LAO substrate are practically identical, demonstrating fully strained films. Contraction of the out-of plane lattice parameter is seen for thinner films. As will be shown, this is presumably caused by a decrease in the concentration of O vacancies with a corresponding increase in $Ni^{3+}$ cations, which have a smaller ionic radius than $Ni^{2+}$, thereby causing a contraction. In the case of Si/SNO (Fig. 1(b)), a mixture of binary oxides SmO, NiO, $NiO_2$, $Ni_2O_3$, and $Sm_2O_3$, which are thermodynamically stable phases under the growth conditions, are observed in addition to SNO.

With respect to XPS measurements, we focus on the Ni $2p_{3/2}$ spectra since the Ni valence state plays an important role in SNO formation. The Ni $2p_{3/2}$ spectra of 80 nm and 200 nm thick SNO on LAO and 120 nm thick SNO on Si are shown in Figs. 2(a), 2(b), and 2(c). On the surface of LAO/SNO, the most intense Ni $2p_{3/2}$ peak is observed at around 855.5 eV and is characteristic of $Ni^{2+}$ and $Ni^{3+}$ ions.[9] However, this peak shifts to around 854.5 eV on Si, which is associated with $Ni^{2+}$. This result is consistent with XRD measurements of Si/SNO since a homogeneous phase of SNO, which would contain mostly $Ni^{3+}$, cannot be obtained due to the substrate-film lattice mismatch. In addition, it is observed that the $Ni^{3+}/Ni^{2+}$ ratio is higher in the thinner LAO/SNO film (~75 %) compared to the thicker film (~54 %). This difference agrees well with the out-of-plane lattice parameter contraction of the 80 nm film since $Ni^{3+}$ has a smaller ionic radius than $Ni^{2+}$. The relatively low $Ni^{3+}$ concentration is likely due to not annealing in high pressure $O_2$, as stated above.

Resistance as a function of temperature was measured on both LAO/SNO films in-plane. A representative *R-T* plot is shown for the 80 nm film in Fig. 3(a). In the first heating-cooling cycle, an MIT is observed at ~85 °C on heating, but on cooling, *R* monotonically increases without signature of MIT. In subsequent cycles, heating curves follow the previous cooling curve, indicating that the $R$ increase upon cooling is related to a permanent change in the film. Furthermore, in subsequent cycles, $T_{MIT}$ shifts to higher temperature (~155 °C in the 4$^{th}$ cycle) and the resistance ratio $\Delta R$ ($\equiv R(25\ °C)/R(T_{MIT})$) increases from 1.02 to 1.25. The 200 nm film shows similar qualitative behavior as the 80 nm film, although the initial values of $T_{MIT}$ (~165 °C) and $\Delta R$ (1.46) are significantly higher. In both films, the initial value of $\Delta R$ is about one order of magnitude smaller as compared to published values.[10] This is consistent with the oxygen non-stoichiometry in our films, observed by XPS, which tends to flatten $R$-$T$ curves by introducing gap states that decrease (increase) resistivity in the semiconducting (metallic) phase.[11] The deviation in $T_{MIT}$ between both samples and as compared to literature values (~110-130 °C)[6,10] may also be caused by oxygen non-stoichiometry with the additional influence of epitaxial strain, which is known to affect the Ni$^{3+}$ valence state and thereby modify $T_{MIT}$ in LnNiO$_3$.[12] Since the LAO/SNO films are compressively strained, $T_{MIT}$ should be lower than the bulk value of 130 °C, which is the case here only for the 80 nm film. However, the higher oxygen non-stoichiometry in the 200 nm film can further increase $T_{MIT}$,[13] resulting in the initial value observed here that is greater than the bulk value.

The $R$-$T$ curves are unusual in the monotonic increase in $R$ upon cooling and the shift of $T_{MIT}$ in successive cycles. For all films, if the temperature is not increased sufficiently high as to observe an MIT, $R$ does not increase significantly upon cooling. This suggests that the MIT causes an irreversible change in the films. In SNO, the cooling curve has been observed to follow the heating curve with hysteresis of a few °C.[6] It has been previously shown with the bulk form of NdNiO$_{3-\delta}$ that with increasing $\delta$, $T_{MIT}$ shifts to higher temperature and the overall resistivity increases.[13] The similarity with results here suggests that our SNO films may lose oxygen after the MIT. Indeed, XRD and XPS measurements before and after several *R-T* cycles (not shown here) show that the films are still single phase after heating but that there is a decrease in the $Ni^{3+}/Ni^{2+}$ ratio, indicative of increasing oxygen non-stoichiometry. Lower $Ni^{3+}/Ni^{2+}$ ratio in the films should lead to more semiconducting behavior.[10] Note that the thermal expansion coefficient of SNO is larger in the metallic state than in the insulating state.[2] Thus, it is possible that perhaps due to low initial oxygen content in our films, the $Ni^{2+}$ state, with larger ionic radius, becomes thermodynamically preferred over the $Ni^{3+}$ state, resulting in oxygen loss. This increase in *R* upon the MIT may be of interest for memristive applications, as discussed below.

For SNO thin films grown on Si, out-of-plane *R-T* measurements (Fig. 3b) show a phase transition in the 1st cycle at ~185 °C that is shifted to ~215 °C in the 3rd cycle. However, because of the existence of multiple phases in Si/SNO as seen by XRD, we note that the MIT observed may not be purely due to a phase transition in SNO. Of the various phases observed, $Ni_2O_3$ and SmO are conducting at room temperature while $Sm_2O_3$, $NiO_2$ (in the ground state), and NiO are insulating.[14-18] It is possible that certain phases dominate the resistance at lower temperatures and others at higher temperatures, reminiscent of phase transition characteristics in $VO_x$ films grown on Si that also contain mixed oxide phases.[19] Note that the *R-T* curves demonstrate similar behavior with cycling as the in-plane measurements in LAO/SNO films. This suggests that the *R-T* measurements on Si/SNO still have some contribution from SNO despite the presence of additional oxide phases.

The change in *R* upon heating above $T_{MIT}$ in our SNO films lasts for weeks and is seemingly permanent. Such behavior is suggestive of memristive systems in which the resistance state of the system is a dynamic function of its past states.[20] Here, memristive behavior in LAO/SNO films is demonstrated by observing changes in *R* after application of 10 V square pulses of 1 s pulse width. $R$ as a function of number of pulses at various sample temperatures is plotted in Fig. 3(c) for an 80 nm LAO/SNO film. At all three temperatures, $R$ clearly increases with each pulse application. After eight $V$ pulses, $R$ saturates at 25 °C but not at 65 °C and 85 °C, which are closer to $T_{MIT}$. The mechanism driving the resistance increase is likely related to the loss of oxygen upon heating as observed in $R$-$T$. A $V$ pulse causes current to flow through the device, which, through Joule heating, can locally induce an MIT in an identical way as external heating. However, it is also possible that the current density during the $V$ pulse is sufficiently high as to effectively dope the channel and drive the MIT as seen in $VO_2$.[21] In either case, it seems as though applying a $V$ pulse is analogous to heating the film to a fixed temperature $T_o$ above $T_{MIT}$, which increases $R$ and shifts $T_{MIT}$ to a higher temperature $T_{MIT}^*$. With each successive $V$ pulse, the difference between $T_o$ and $T_{MIT}^*$ decreases and eventually the resistance increase saturates. At elevated temperature, a $V$ pulse will have greater effect than at room temperature and more pulses are needed before saturation. Additionally, our LAO/SNO device also exhibits several of the general properties derived of memrisitive systems,[20] such as passivity ($R > 0$), no energy discharge ($V = 0$ for $I = 0$ and vice versa), and nonlinear DC resistance ($dV/dI \neq$ constant, under low frequency operation). These preliminary results are suggestive of memristive behavior and also imply that it may be possible to induce the phase transition of SNO with an applied voltage.

In summary, SNO films were grown on LAO (001) and Si (100) by RF magnetron sputtering. On LAO substrates, SNO films grew fully strained in a single phase. Electrical measurements showed an MIT with increasing $T_{MIT}$ on each successive heating-cooling cycle and irreversible resistance increase after cooling. SNO films deposited under similar conditions on Si contain multiple phases of Sm and Ni oxides. Out-of-plane $R$-$T$ measurements on Si/SNO reveal an MIT in vertical devices. With voltage pulse application, electrically driven memristive behavior in LAO/SNO films is observed.

The authors acknowledge ARO MURI (W911-NF-09-1-0398) and the Focus Center Research Program in the Materials Structures and Devices Focus Center for financial support.

**Fig. 1:** $2\theta$-$\omega$ scans of (a) LAO/SNO (Inset: $\varphi$ scan of LAO/SNO) and (b) Si/SNO as deposited. Reciprocal space mapping of the 103 asymmetric reflection of (c) 80 nm and (d) 200 nm LAO/SNO.

**Fig. 2:** Ni $2p_{3/2}$ XPS spectra of (a) 80 nm and (b) 200 nm LAO/SNO and (c) 120 nm Si/SNO films. The peaks associated with Ni, $Ni^{2+}$, and $Ni^{3+}$ are indicated. The relative concentration of $Ni^{3+}$ and $Ni^{2+}$ was determined from the area of core level peaks in the region of Ni $2p_{3/2}$ (~851-858.5 eV) after Shirley background subtraction and assuming a Voigt line shape for the peaks.

**Fig. 3:** *R-T* measurements of (a) 80 nm LAO/SNO and (b) 120 nm Si/SNO normalized to *R*(25 °C) of 1st cycles. Insets: Raw *R-T* data for respective 3rd cycles. (c) Resistance after applying 10 V, 1 s square pulses to in-plane 80 nm LAO/SNO device.

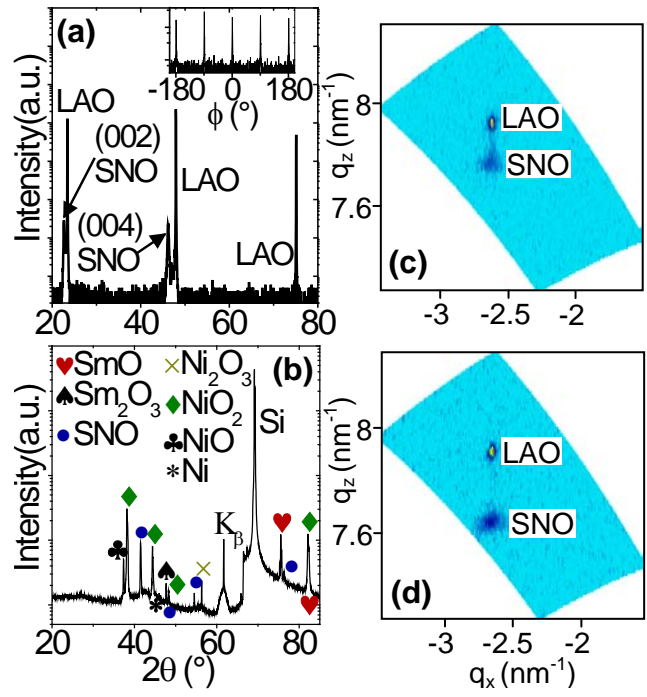

Fig. 1

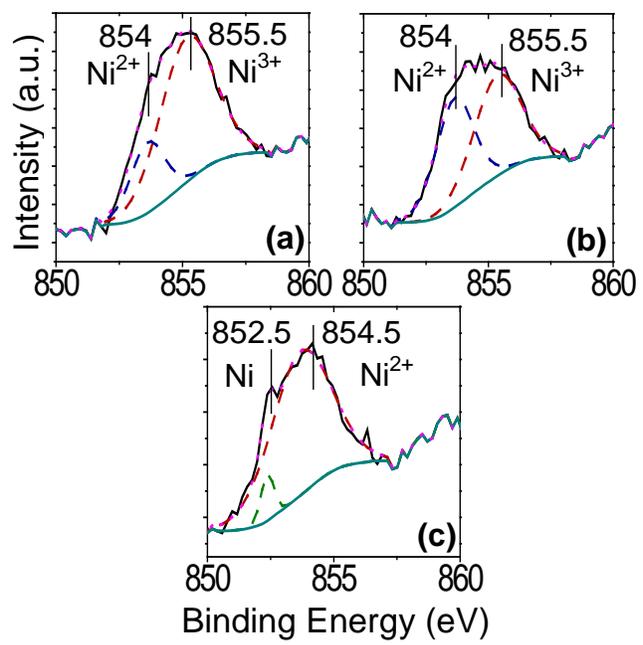

Fig. 2

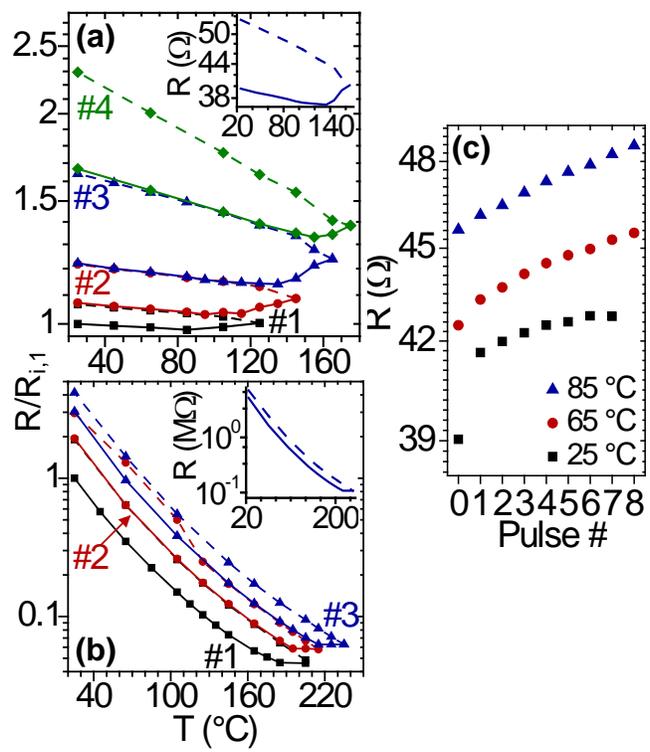

Fig. 3